\documentclass[journal]{vgtc}                     

\onlineid{0}

\vgtccategory{Research}

\vgtcpapertype{please specify}

\title{\tool{}: Exploring Multimodal Literature Foraging Strategies in Immersive Sensemaking}

\author{%
  \authororcid{Haoyang Yang}{0000-0002-0566-0169},
  \authororcid{Elliott H. Faa}{0009-0002-8698-0961},
  \authororcid{Weijian Liu}{0009-0004-4771-2539},
  \authororcid{Shunan Guo}{0000-0001-5355-8399},
  \authororcid{Duen Horng Chau}{0000-0001-9824-3323},
  and 
  \authororcid{Yalong Yang}{0000-0001-9414-9911}
}

\authorfooter{
  \item
  	Haoyang Yang, Elliott H. Faa, Weijian Liu, Duen Horng Chau, and Yalong Yang are with Georgia Tech. 
    E-mail: \{alexanderyang, efaa3, wliu430, polo, yalong.yang\}@gatech.edu
  \item
  	Shunan Guo is with Adobe Research.
  	E-mail: sguo@adobe.com
}

\abstract{Exploring and comprehending relevant academic literature is a vital yet challenging task for researchers, especially given the rapid expansion in research publications. This task fundamentally involves sensemaking—interpreting complex, scattered information sources to build understanding.
    While emerging immersive analytics tools have shown cognitive benefits like enhanced spatial memory and reduced mental load, they predominantly focus on \textit{information synthesis} (e.g., organizing known documents). 
    In contrast, the equally important \textit{information foraging} phase—discovering and gathering relevant literature—remains underexplored within immersive environments, hindering a complete sensemaking workflow. 
    To bridge this gap, we introduce \tool{}, an interactive literature exploration tool designed to facilitate information foraging of research literature within an immersive sensemaking workflow using network-based visualizations and multimodal interactions.
    \added{Developed with WebXR and informed by a formative study with researchers,} \tool{} supports exploration guidance, spatial organization, and seamless transition through a 3D literature network. \added{An observational user study with 15 researchers} demonstrated \tool{}'s effectiveness in supporting fluid foraging strategies and spatial sensemaking through its multimodal interface.
} 


\teaser{
  \centering
  \includegraphics[width=\linewidth]{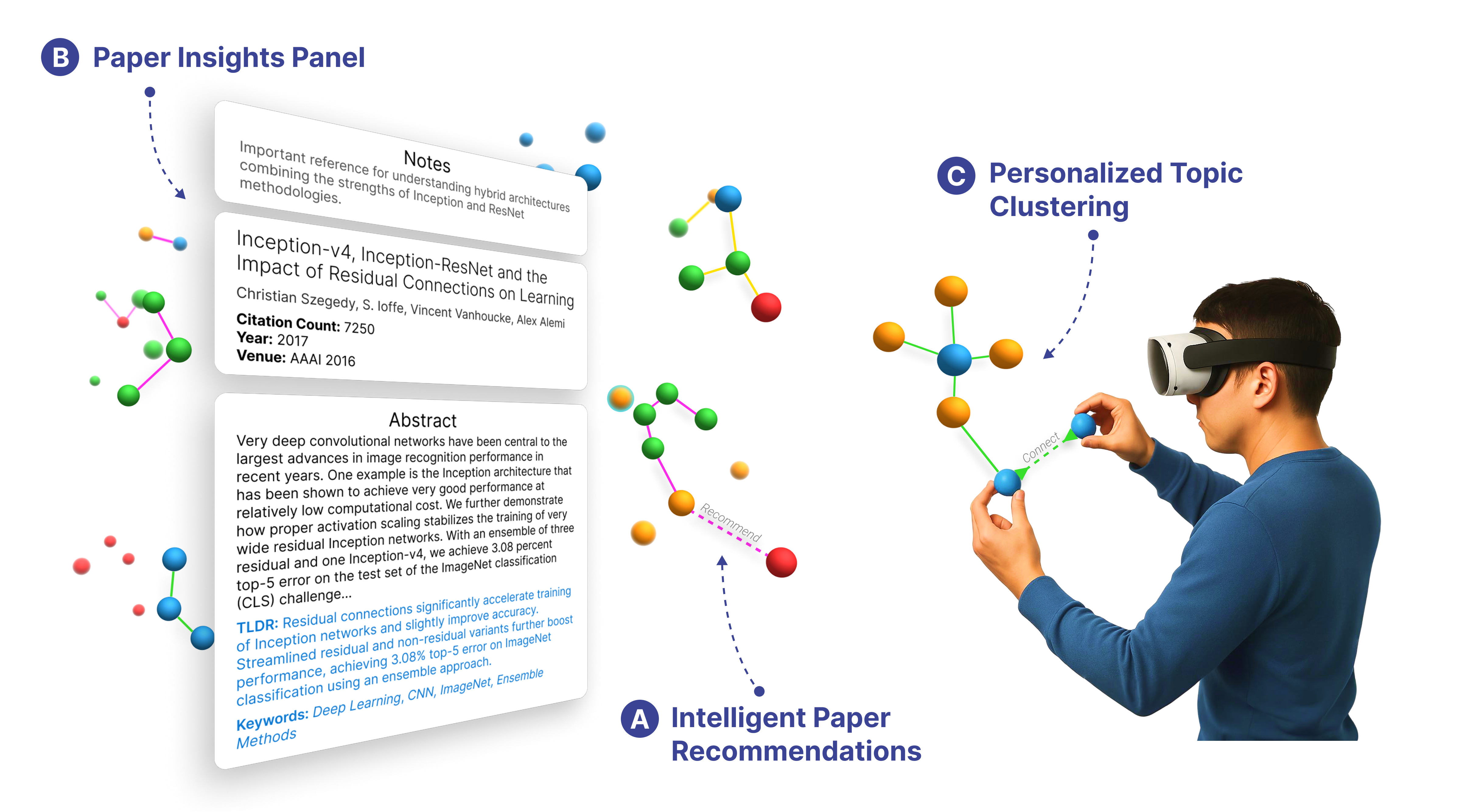}
   \vspace{-2em}
  \caption{\tool{} supports multimodal interactions to enable researchers to forage and visualize paper nodes within an immersive literature network graph in VR.
\textbf{(A) Intelligent Paper Recommendations} help forage relevant literature via paper connections, such as citations, co-authorship, and thematic similarities.
\textbf{(B) Paper Insights Panel} provides paper summaries and conceptual highlights while supporting user-authored annotations to capture reflections and emerging ideas.
\textbf{(C) Personalized Topic Clustering} automatically groups papers by theme, with gesture controls allowing flexible, user-driven graph customization.}
  \label{fig:teaser}
}

\PassOptionsToPackage{svgnames}{xcolor}
\usepackage{xcolor}
\PassOptionsToPackage{colorlinks}{hyperref} 

\usepackage{enumitem}  
\usepackage{wrapfig}
\usepackage[defaultcolor=black]{changes}

\setitemize{noitemsep,topsep=0pt,parsep=0pt,partopsep=0pt}

\definecolor{agreen}{RGB}{74, 198, 148}
\definecolor{purple}{RGB}{158, 62, 177}  
\definecolor{darkpurple}{RGB}{170, 70, 210}
\definecolor{aqua}{RGB}{87, 180, 181}
\definecolor{lightblue}{RGB}{72, 123, 232}
\definecolor{hotpink}{RGB}{255, 83, 115}
\definecolor{teal}{RGB}{90, 200, 250}
\definecolor{linkColor}{RGB}{0, 128, 229}
\definecolor{lightgreen}{RGB}{33, 222, 128}
\definecolor{almostBlack}{RGB}{60,60,60}
\definecolor{red}{RGB}{255, 0, 0}
\definecolor{green}{RGB}{0, 128, 0}
\definecolor{yellow}{RGB}{255, 192, 0}
\definecolor{cyan}{RGB}{0, 255, 255}
\definecolor{lightgray}{gray}{0.95}
\definecolor{grayborder}{gray}{0.5}
\definecolor{gray}{gray}{0.75}
\definecolor{orange}{RGB}{236, 107, 44}
\definecolor{lightorange}{RGB}{255, 223, 186}
\definecolor{lightpurple}{RGB}{202,58,126}
\definecolor{benign_purple}{RGB}{150,150,150}
\definecolor{adv_orange}{RGB}{212,64,57}

\newcommand{\tool}{\textsc{LitForager}\xspace{}}

\AtBeginDocument{%
  \hypersetup{
    colorlinks,
    linkcolor=niceblue,
    citecolor=niceblue,
    urlcolor=niceblue,
    filecolor=niceblue
  }%
  \crefname{figure}{fig.}{fig.}
    \Crefname{figure}{Fig.}{Fig.}
    \crefname{equation}{eq.}{eq.}
    \Crefname{equation}{Eq.}{Eq.}
    \crefname{section}{\S}{\S}
    \creflabelformat{equation}{#2\textup{#1}#3}
}

\definecolor{niceblue}{RGB}{56, 116, 203}
\newcommand{\figpart}[1]{\textcolor{niceblue}{#1}}
\graphicspath{{figs/}{figures/}{pictures/}{images/}{./}} 

\usepackage{tabu}                      
\usepackage{booktabs}                  
\usepackage{lipsum}                    
\usepackage{mwe}                       

\usepackage{mathptmx}                  

\begin{document}


\firstsection{Introduction}
\maketitle
Literature review is a crucial part of virtually every research endeavor, yet staying on top of the influx of new papers has become increasingly challenging~\cite{landhuis_scientific_2016}. Each year, millions of new research articles are published~\cite{knoth_core_2023, johnson2018stm, ghasemi_scientific_2022}, making it difficult for researchers to discover and digest the latest relevant information.

Literature review is fundamentally a sensemaking activity that involves both information foraging and synthesis. 
Pirolli and Card's sensemaking model~\cite{pirolli1999information,pirolli2005sensemaking} formalizes this process as two interlinked loops: an \textit{information foraging} loop of searching, filtering, and collecting pertinent data sources, followed by an \textit{information synthesis} loop of analyzing and synthesizing insights from those sources. 
Effective sensemaking is an iterative process between the two phases, but most prior research focused on only one phase rather than integrating both, leaving the full iterative workflow underexplored~\cite{endert_state_2017}. 
This gap is particularly evident in the emerging field of immersive analytics. 
Despite the recognized cognitive benefits of immersive environments---such as enhanced spatial memory and reduced cognitive load---current systems are mainly designed to support synthesis tasks, such as helping users organize pre-existing document collections or visualize data in 3D~\cite{lisle_evaluating_2020, lisle_sensemaking_2021, davidson_exploring_2023}. Support for the upstream task of information foraging within the immersive space remains limited, reducing the potential of immersive technologies to facilitate the complete, end-to-end sensemaking workflow.

To address this gap and explore the potential of immersive sensemaking with integrated foraging support, we introduce \tool{}, an immersive system explicitly designed to streamline information foraging during literature exploration. 
To shape the design of \tool{}, we conducted a formative study with expert interviews to identify researchers' goals, needs, and challenges during literature review. 
Insights from this study highlighted the need for effective \textbf{exploration guidance}, flexible \textbf{spatial organization}, and \textbf{seamless transition} between foraging and synthesis, which became the core design goals for \tool{}.

To provide \textbf{exploration guidance}, \tool{} presents literature as an interactive 3D network graph, visualizing relationships like citations, authorship, and thematic similarities. Users can dynamically expand this network using integrated recommendations from Semantic Scholar~\cite{Ammar2018ConstructionOT, Lo2020S2ORCTS}, and rapidly comprehend content with the aid of LLM-powered summaries and keyword extraction. 
\tool{} supports flexible \textbf{spatial organization}---users can freely position papers within the 3D space or apply automatic clustering by topic similarity. This allows them to leverage spatial memory and externalize their mental model. 
To enable \textbf{seamless transition} between foraging and synthesizing in the iterative workflow, the system integrates multimodal input, combining intuitive hand gestures for direct manipulation with voice commands for efficient control over exploration, annotation, and workspace management. 
\added{Implemented with Babylon.js~\cite{babylonjs} and WebXR~\cite{webxr} and guided by user requirements identified in our formative study, \tool{} provides an embodied, web-based environment tailored for effective literature foraging. We evaluated \tool{} through an observational user study with 15 researchers who engaged in immersive literature discovery tasks to assess the system’s usability and interaction design. }

In summary, our work has the following primary contributions: 
\begin{itemize}[topsep=1pt, itemsep=0mm, parsep=3pt, leftmargin=9pt]
    \item \textbf{A formative study with six academic researchers} that identifies critical user needs, task breakdown, and pain points faced during literature discovery and sensemaking. The findings informed the key requirements and design considerations necessary to effectively support immersive literature exploration.
    \item \textbf{\tool{}, a fully developed, open-source\footnote{\tool{}'s source code is publicly available at \url{https://github.com/AlexanderHYang/LitForager}.} immersive system explicitly designed to support literature foraging in the sensemaking workflow}. \tool{} integrates literature network visualization, spatial organization, and multimodal interactions, allowing users to explore and comprehend complex relationships within the literature and enhance cognitive engagement and memory.

    \item \added{\textbf{An observational study with 15 academic researchers} that provides insights into how \tool{} can support literature foraging and sensemaking workflow}. Our findings suggest that spatial organization features promote externalized thinking, while multimodal interaction enables fluid exploration.
\end{itemize}
\section{Related Work}

\noindent\textbf{Immersive Sensemaking.}
Early cognitive models of sensemaking, such as the framework by Pirolli and Card, describe iterative \textit{foraging} and \textit{synthesis} loops through which analysts gather information and develop hypotheses~\cite{russell_cost_1993, pirolli1999information, pirolli2005sensemaking}. 
\textit{Foraging} involves searching and filtering information into schemas, while \textit{synthesis} entails refining a mental model based on these schemas. Building on this foundation, later research has highlighted the importance of externalizing thought processes to amplify cognition. 
Notably, the Space to Think system~\cite{andrews2011information} demonstrated that using a large display allowed researchers to spatially arrange information and offload memory, significantly improving sensemaking performance. 

Leveraging space as an extension of cognition is central to the emergence of immersive analytics. Immersive systems provide an expansive 3D canvas and engage the user’s body through movement and gesture, enabling spatial cognition and embodied interaction for complex analysis~\cite{tong2025exploring,tong2023towards}. Drawing inspiration from Space to Think, Lisle et al. introduced Immersive Space to Think (IST), a VR-based sensemaking environment for document analysis~\cite{lisle_sensemaking_2021}, \added{and later compared the cognitive effects of traditional 2D and immersive displays~\cite{lisle_spaces_2023}, showing how immersive environments leverage spatial memory and externalization to reduce context-switching costs and enhance sensemaking. Tahmid et al.~\cite{tahmid_evaluating_2023, tahmid_enhancing_2025} extended IST by integrating gaze-driven multimodal interactions and recommendation cues to support dynamic information discovery and relevance prediction.} However, both IST and its extensions rely on pre-selected
document corpora and offer limited support for integrating newly discovered materials into the workflow, restricting their adaptability for real-world, iterative foraging tasks across evolving datasets.


Our tool addresses these limitations by envisioning a complete sensemaking workflow with built-in discovery and recommendation capabilities. In contrast to prior systems that required manually pre-loaded document sets, \tool{} enables users to fluidly discover new papers and insert them into their analysis environment on the fly, bringing the traditionally separate discovery phase into the same immersive workflow.

\smallskip 
\noindent\textbf{Literature Discovery.}
Literature discovery is the process of exploring academic literature to uncover insights, patterns, or connections that are not explicitly stated~\cite{smalheiser_literature-based_2012, smalheiser_rediscovering_2017}.  
It differs from the standard literature search in that it is a more exploratory process rather than a straightforward goal-oriented query.
In a standard search, a researcher typically knows what to look for: specific keywords, authors, or papers. Literature discovery, on the other hand, can be more beneficial when the researcher does not have a precise target in mind; instead, the goal is to explore a research area and uncover unexpected yet relevant connections. 

Many interactive systems support literature discovery by helping researchers navigate relationships between works. Kang et al.~\cite{kang_comlittee_2023} focus on author networks, allowing users to form a ``committee'' of key authors and expand recommendations via co-authorship and citation links. PaperQuest~\cite{ponsard_paperquest_2016} visualizes citation networks and applies relevance algorithms to guide exploration. Palani et al.~\cite{palani_relatedly_2023} extract and compile related work sections to highlight key prior studies and recurring themes. PUREsuggest~\cite{beck_puresuggest_2025} combines citation-based recommendations with interactive visual exploration, allowing users to guide the process using seed papers and keywords. Building on these approaches, our system supports literature discovery by enabling researchers to explore, connect, and uncover insights through immersive foraging.

\smallskip 
\noindent\textbf{Multimodal and Embodied Interaction in VR.}
Prior research in multimodal interaction shows that combining input modalities enhances usability across workflows. For example, Saktheeswaran et al.~\cite{saktheeswaran_touch_2020} explored how touch and speech support network analysis, while Data@Hand~\cite{kim_datahand_2021} enabled personal data exploration on smartphones using the same modalities. Reinders et al.~\cite{reinders_when_2025} paired conversational agents with tactile displays to improve accessibility for blind users. \added{Luo et al.~\cite{luo_documents_2025} demonstrated how bimanual gestures and spatial manipulation in AR support intuitive document arrangement.} Collectively, these studies highlight the versatility of multimodal interfaces in improving user experience.

In immersive environments, speech input enables natural command execution and querying without typing. MEinVR~\cite{yuan_meinvr_2022} combined voice and controller input to navigate 3D molecular structures more intuitively. Multimodal interactions are also central in commercial systems: the Microsoft HoloLens 2 blends eye gaze, hand gestures, and voice for seamless interaction~\cite{multimodal_microsoft}, reducing reliance on abstract UI elements like WIMP menus.

Immersive analytics also benefits from embodied interaction and spatial layout, which offload cognitive effort through physical action and spatial memory~\cite{cordeil_imaxes_2017, huang2023embodied, yang_tilt_2021, in_evaluating_2024, zhu_compositingvis_2024}. By allowing analysts to move and organize content in 3D space, immersive systems act as external memory aids~\cite{skarbez_immersive_2019, lisle_sensemaking_2021}, minimizing the need for internal recall or auxiliary note-taking~\cite{hutchins1995cognition, norman2013design, kirsh1995intelligent, yousif_using_2021}. \added{Lisle et al.~\cite{lisle_different_2023} and Luo et al.~\cite{luo_where_2022} demonstrated that both VR and AR environments enable such memory externalization by allowing documents to be anchored relative to virtual objects or physical structures.} Similarly, Andrews et al.~\cite{andrews2011information} found that users working with large displays leveraged spatial positioning to track document relationships and maintain context over time. Such embodied interaction reduces memory load and mirrors real-world behaviors—like grouping items or glancing around to recall a location—making complex tools more intuitive and learnable.

\section{Formative Study}
\label{3}
\begin{figure}
    \centering
    \includegraphics[width=\linewidth]{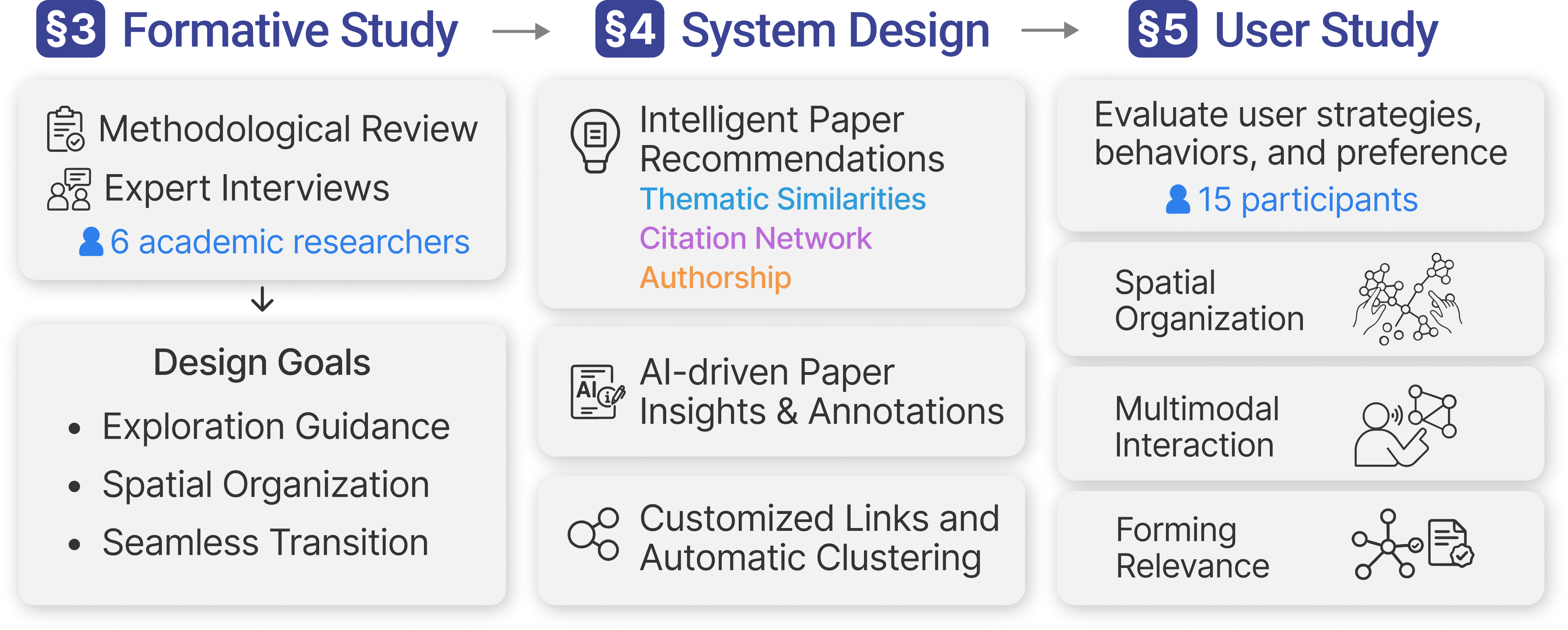}
    \caption{Overview of our research workflow. We conducted a formative study~(\autoref{3}) to identify design goals for immersive literature foraging. Based on these goals, we developed a system~(\autoref{4}) with interactive foraging features. We then evaluated the system~(\autoref{5}) through a user study with 15 participants, focusing on spatial organization, multimodal interaction, and relevance formation.}
    \label{fig:pipeline}

\end{figure}
\added{In this section, we present a formative study based on semi-structured expert interviews that investigates how academic researchers conduct literature reviews.} We examine their primary tasks, objectives, and pain points in the process, which informed the design goals of \tool{}.

\subsection{Procedures}


\noindent
\textbf{Semi-Structured Expert Interviews.} 
\added{We reviewed influential educational resources and guidelines on literature reviews, including those from university libraries and educational institutions~\cite{ucsd_lit_review_guideline, brown_lit_review_guideline, purdue_lit_review_guideline, pare_chapter_2017} to ground our understanding of discovery behaviors, strategies for evaluating relevance, and synthesis practices. Booth et al.’s systematic framework~\cite{booth_systematic_2016} further guided the segmentation of the literature review process into distinct stages (searching, organizing, and synthesizing), which shaped the structure of our study procedure. Together, these sources influenced the design of our open-ended interview questions, including how we framed prompts around identifying starting points, expanding search strategies, assessing importance, and integrating insights.}

We conducted semi-structured interviews with 6 academic researchers \added{(R1–6; 2 female, 4 male; ages 20-25)}, including five PhD students and one master’s student, all with extensive literature review experience. 
Prior to the interviews, participants completed a brief survey detailing their research experience, literature review frequency, and objectives, providing additional context for discussion. The interview questions were organized according to distinct stages of the literature review process, with greater emphasis placed on the phases of literature searching and discovery, reflecting our focus on information-foraging behaviors.
Specifically, we asked participants about their strategies for identifying starting points, expanding their search, organizing literature, evaluating relevance, and synthesizing insights when conducting a literature review.
We iteratively refined the interview protocol through two pilot sessions. Each interview lasted approximately 45–60 minutes and was conducted in person. 

\smallskip 
\noindent
\textbf{Data Collection and Analysis.} All interviews were recorded (with consent) and transcribed for analysis. We collected additional notes on any specific tools or techniques participants mentioned (e.g., using reference managers, annotation practices, or ad-hoc physical organization of papers). Using Braun and Clarke's reflective approach~\cite{braun_reflecting_2019, byrne_worked_2022}, we performed a thematic analysis on the transcripts to extract recurrent themes regarding user tasks, motivations, pain points, and strategies. Three authors independently coded one of the transcripts to develop an initial schema. The thematic coding was refined through discussion and then applied to re-code all six transcripts. Consistency and saturation were reached by the fifth and sixth interviews, as no substantially new themes emerged.

\subsection{Findings \& Insights}
We present formative study findings that reveal a multi-dimensional view of researchers’ practices and challenges in the literature review process.

\subsubsection{Context, Data Needs, and Practices}
\label{3.2.1}
\begin{figure}
    \centering
    \includegraphics[width=\linewidth]{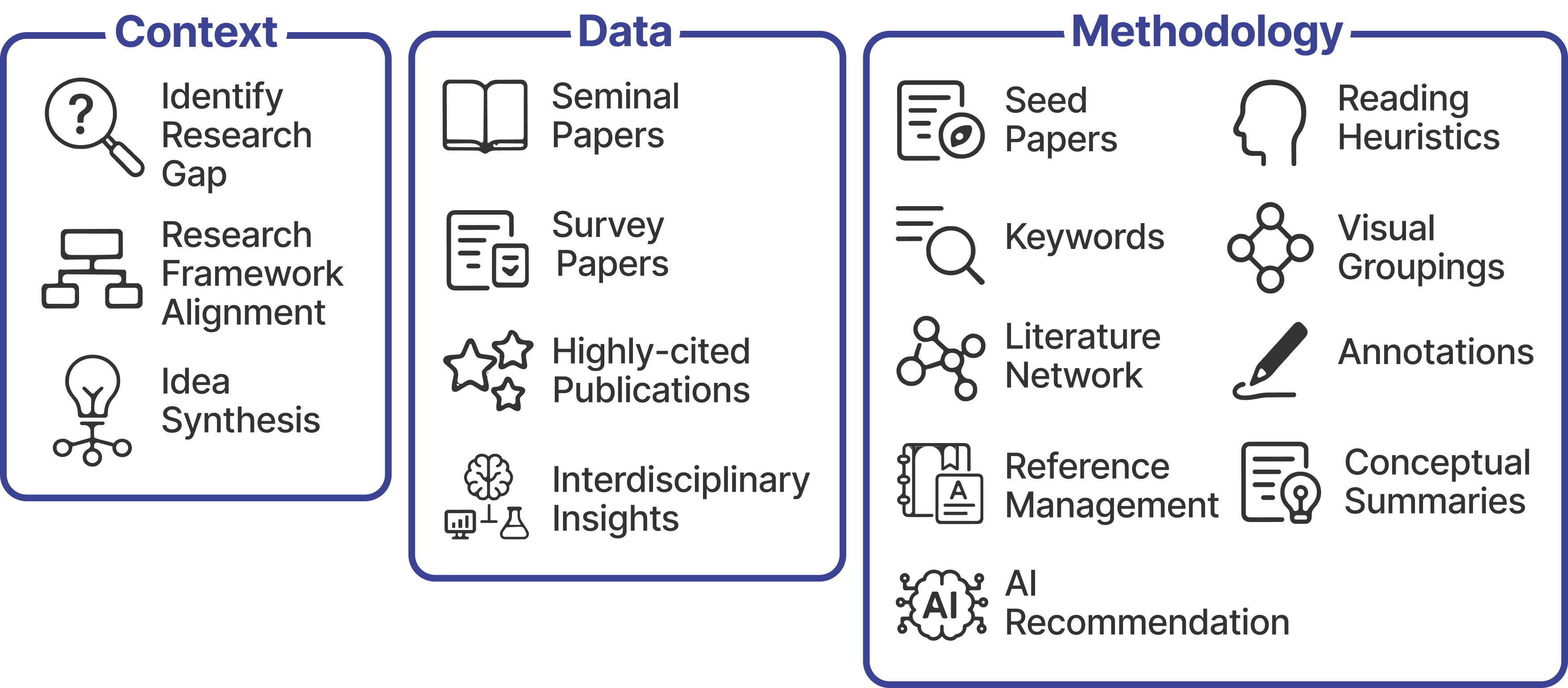}
    \caption{Three interrelated dimensions of literature review practices identified from expert interviews.}
    \label{fig:theme-categories}
\end{figure}

From the expert interviews,
we identified both distinctive strategies and common themes across three interrelated dimensions~(\autoref{fig:theme-categories}): \textbf{Context} (\textit{When} and \textit{Why} researchers conduct literature reviews), \textbf{Data} (\textit{What} information researchers seek), and \textbf{Methodology} (\textit{How} they source, evaluate, and organize the information)~:

\smallskip 
\noindent
\textbf{Context.}
Researchers reported engaging in literature reviews at various points throughout their projects. During the early stages of a research project (initiation and ideation), researchers conduct literature reviews primarily to \textit{\textbf{identify research gap}}, which helps define the project’s scope and refine research questions, \textit{``\dots to do something really novel, I have to find papers that I can build on''} (R2). Later, in the mid-project phase (development and problem-solving), researchers revisit the literature review primarily for research \textit{\textbf{framework alignment}}, positioning their findings within broader scholarly conversations and updating evolving theoretical frameworks. This alignment ensures their work remains relevant to current trends and clearly demonstrates the novelty of their contributions. In the paper writing phase, the literature review process shifts towards \textit{\textbf{idea synthesis}}. As researchers gather more insights from additional readings, they organically develop subsections within their related work, structuring emerging sub-concepts to frame their findings within a broader scholarly conversation, \textit{``As I read more, I finalize these subconcepts into the subsections of my related work''} (R4).

\smallskip 
\noindent
\textbf{Data.}
The interview data reveal that researchers look for a variety of information types during the literature review process. Many researchers indicated that they build a comprehensive understanding of their field by starting with \textit{\textbf{seminal papers}} or \textit{\textbf{survey papers}}, which encapsulate foundational concepts and frameworks, \textit{``If I’m completely new to a field, I start with literature surveys or seminal papers''} (R3). There is also a significant focus on emerging trends and identifying gaps: many researchers sift through \textit{\textbf{highly-cited publications}} such as conference proceedings or journal articles, \textit{``I lean more on quantitative metrics like citations, but also on the reputation of the venue''} (R1), to capture current developments and highlight areas where further inquiry is needed. Additionally, \textit{\textbf{interdisciplinary insights}} also play a vital role. Researchers noted that sometimes, by exploring literature from other fields, they uncover novel perspectives, \textit{``I notice that certain techniques used elsewhere might be applicable—even if the fields seem disconnected''} (R6) and \textit{``I came across a recommended study about policy and governance aspects of the field that wasn’t originally on my radar''} (R3). This diverse set of data ensures that their work is well-grounded in established knowledge while simultaneously aligning it with state-of-the-art developments in emerging research.

\smallskip 
\noindent
\textbf{Methodology.}
In tandem with the information they seek, researchers have mentioned various methodologies for sourcing, evaluating, and organizing the literature. They usually start out with a few \textit{\textbf{seed papers}}, often recommendations from peers and guidance from experts or advisors, \textit{``Often, I begin my literature research with a few `seed' papers that my peers or my advisor have given me or that I happened to catch during a conference talk''} (R3).  When sourcing relevant studies, they leverage targeted searches using specific \textit{\textbf{keywords}}, often complemented by various types of \textit{\textbf{literature network}} to trace connections across the research landscape, such as citation network, reference network, or author network. Additionally, \textit{\textbf{AI recommendation}} systems further enhance the discovery process by efficiently identifying relevant studies, particularly interdisciplinary works that span diverse academic fields, \textit{``I also use tools like Semantic Scholar’s AI-powered recommendation system to suggest new papers based on the one I’m reading''} (R2).
Once potential sources are identified, researchers often use \textit{\textbf{reading heuristics}} by quickly scanning abstracts, introductions, and conclusions to determine relevance, \textit{``I typically start by skimming the abstract, introduction, and conclusion''} (R3), followed by a more detailed evaluation of the research methodology.
To synthesize the gathered insights, some researchers mentioned creating \textbf{\textit{visual groupings}}, \textit{``I think a mental map is good, especially a tree or network structure \dots''} (R5), \textit{\textbf{annotations}}, and \textit{\textbf{conceptual summaries}} that help identify common themes and discrepancies across studies. 
Finally, effective organization is achieved through the use of \textit{\textbf{reference management}} systems, such as Zotero or Mendeley. 

\subsubsection{Pain Points in the Literature Review Process}

\noindent
\textbf{Information Overload.}
A recurring concern is the sheer scale of available literature. Many mentioned the risk of missing critical studies amid the vast number of publications. This overwhelming volume not only complicates the discovery process but also increases the likelihood of redundancy or overlooked work, \textit{``It’s hard to be completely comprehensive \dots Not finding literature in a particular niche doesn’t necessarily mean there isn’t any''} (R1). Consequently, researchers highlighted the need for effective exploration guidance that can navigate large datasets and surface key papers, ensuring that important contributions are not lost in the noise. Evaluating the credibility and relevance of each source is another labor-intensive aspect of the literature review process. Researchers reported the challenge of having to read through a large number of papers, emphasizing the necessity for reading guidance to streamline the evaluation process.

\smallskip 
\noindent
\textbf{Interdisciplinary Blind Spots.}
Many researchers mentioned the concern that, despite their best efforts, critical literature might be inadvertently overlooked. This challenge is particularly acute in interdisciplinary research areas, where seminal studies and innovative methodologies may be scattered across diverse fields and publication venues. Researchers often rely on traditional search strategies that focus on their primary discipline, which can lead to a narrow perspective that misses valuable insights from adjacent or seemingly unrelated domains, \textit{``I wish recognition systems could combine multiple references or keywords simultaneously to yield more interdisciplinary results''} (R3).

\smallskip 
\noindent
\textbf{Difficulty in Synthesis and Organization.}
Despite the availability of digital tools, synthesizing information from a wide range of sources remains a significant challenge. Researchers noted that while reference managers help with storage, they fall short in supporting synthesis. For instance, tools like Zotero typically organize literature as a linear list, which limits their ability to support visualizations or map thematic grouping. While importing literature into dedicated visual mapping tools could be a solution, it adds workflow friction. Participants expressed a need for tools that allow them to visually map connections and structure their ideas, while also integrating the foraging aspect of exploring a large corpus of literature, \textit{``I envision a system where I could offload some cognitive effort by interacting with a dynamic graph that expands into new nodes as I explore ideas further''} (R6). Researchers require a seamless, iterative workflow that enables them to transition effortlessly from discovering and storing relevant studies to synthesizing and constructing a coherent narrative.

\subsection{Design Goals}
\label{3.3}

Drawing from the insights of our formative study and grounded in established HCI and sensemaking principles, we distilled three core design goals (\ref{goal:guidance}-\ref{goal:interaction}) to guide the development of \tool{}. 
These goals directly address the identified researcher pain points and aim to integrate effective foraging support within an immersive sensemaking workflow. 

\begin{enumerate}[topsep=2pt, itemsep=0mm, parsep=3pt, leftmargin=19pt, wide, labelwidth=!, labelindent=0pt, label=\textbf{G\arabic*.}, ref=G\arabic*]
\item \label{goal:guidance} \textbf{Intelligent exploration guidance via recommendations and insights.} The challenge of navigating information overload and evaluating relevance highlights the need for intelligent exploration guidance.  Aligned with Information Foraging Theory~\cite{pirolli1999information}, our design focuses on surfacing relevant and high-quality literature through targeted recommendations and AI-driven insights, helping researchers efficiently navigate large datasets and emerging trends while reducing reliance on exhaustive manual search.

\item \label{goal:organization} \textbf{Flexible spatial organization of the literature network.}
Organizing and synthesizing diverse sources using traditional foraging tools can be cognitively demanding. Leveraging the affordances of immersive environments~\cite{andrews2011information,lisle_sensemaking_2021,in_evaluating_2024,in_this_2024}, \tool{} introduces a 3D literature network that enables researchers to externalize their understanding by spatially arranging papers. Supporting both automatic clustering and customizable layouts, this design aims to help offload cognitive effort and facilitates intuitive exploration of thematic and methodological relationships.

\item \label{goal:interaction} \textbf{Seamless transition between foraging and synthesizing.} Synthesizing information from a vast corpus of literature imposes significant cognitive load, especially when traditional tools separate the processes of finding and organizing information. To address this disconnect, we aim to create a seamless transition between foraging and synthesizing by integrating robust exploration with fluid multimodal interaction. By reducing context switching and interface complexity~\cite{pirolli2005sensemaking}, our goal is to help researchers maintain context and focus throughout the iterative sensemaking process.

\end{enumerate}

\section{\tool{}}
\label{4}


Building on the insights from our formative study~(\autoref{3}) and guided by our core design goals~(\autoref{3.3}), we developed \tool{}, an interactive prototype designed to make literature foraging easier within an immersive sensemaking workflow. 
This section details the core components and features of \tool{}, organized by the literature network visualization, interaction model, and key features enabling intelligent immersive foraging and sensemaking.

\subsection{Literature Network Visualization}
\label{4.1}

\setlength{\columnsep}{5pt}%
\setlength{\intextsep}{0pt}%
\begin{wrapfigure}{R}{0.24\textwidth}
   \vspace{0pt}
   \centering
   \includegraphics[width=0.24\textwidth]{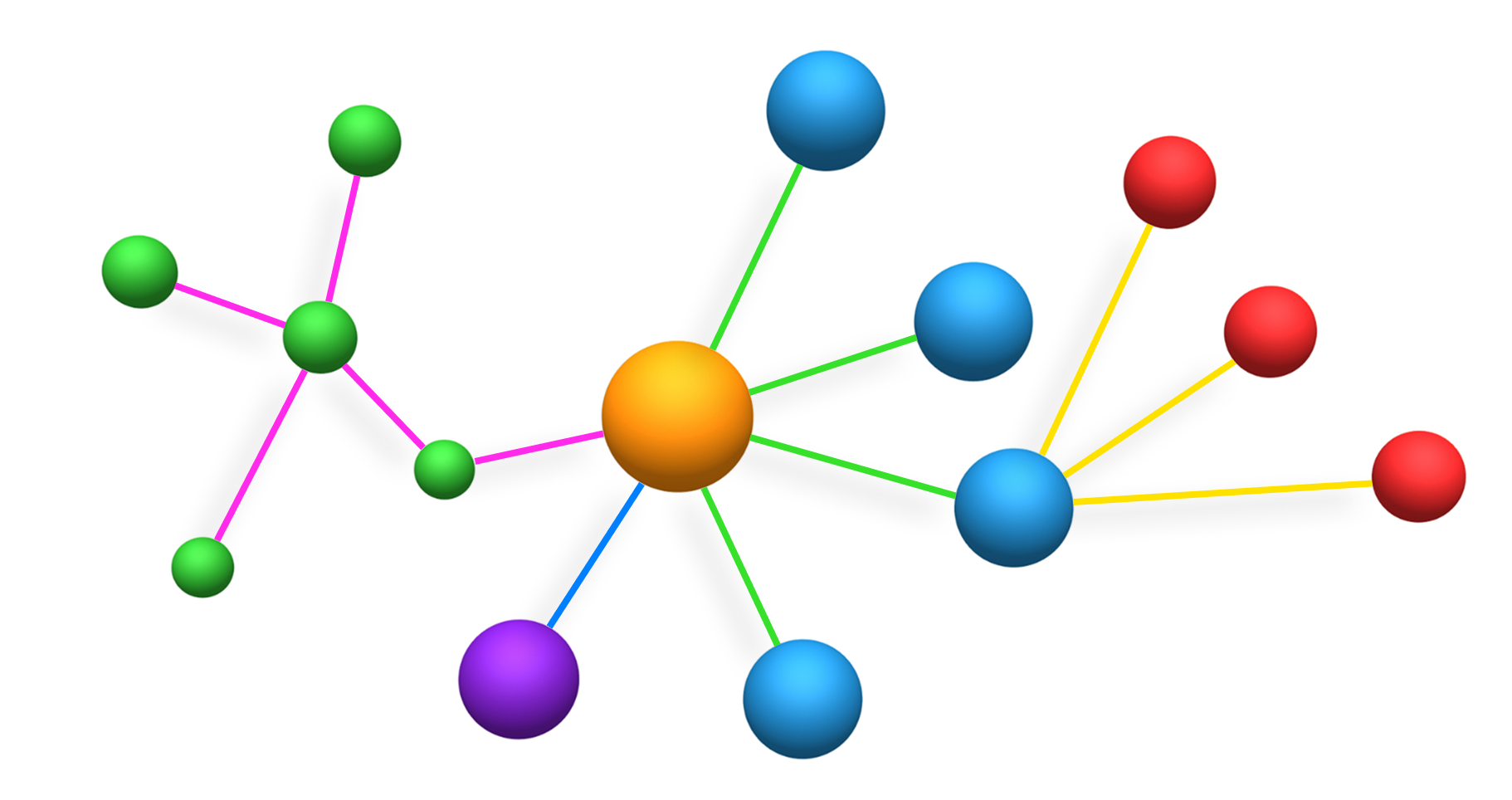}
   \vspace{-10pt}
   \label{fig:lit-network}
\end{wrapfigure}
The core representation within \tool{} is an interactive 3D force-directed graph displaying academic papers as nodes and their relationships using distinct edge colors to convey meaning: white for \textit{thematic similarities}, magenta for \textit{citations \& references}, yellow for \textit{author-centric}, and green for \textit{user-created links}.

Web-based tools like Connected Papers~\cite{ConnectedPapers}, ResearchRabbit~\cite{ResearchRabbit}, and Argo Scholar~\cite{li2022visual} use network visualization to reveal structures, clusters, and pathways within research areas.
Building on this proven approach and research suggesting immersive environments enhance complex network exploration via improved spatial understanding and interaction~\cite{kwon2016study,kotlarek2020study}, \tool{} employs immersive 3D network visualization. \added{To populate the visualization, \tool{} retrieves metadata for requested papers (e.g., authors, abstracts, publication year, and citation relationships) from Semantic Scholar’s scholarly corpus via their public API \cite{Ammar2018ConstructionOT, Lo2020S2ORCTS}.} The visualization, implemented using Babylon.js~\cite{babylonjs} and WebXR~\cite{webxr}, is accessible via web browsers on standard VR headsets.

Initially, the network’s spatial arrangement is determined by a force-directed layout, which automatically clusters related papers in close proximity. \added{To simulate the layout of nodes in 3D space, \tool{} employs d3-force-3d~\cite{d3-force-3d}, a three-dimensional extension of D3’s physical force-directed layout engine~\cite{D3.js}.}
This provides an immediate visual structure that helps users grasp the overall landscape~(\ref{goal:organization}). Rendering this network in VR aims to capitalize on distinct advantages over conventional 2D screens. 
The expansive 3D space allows users to leverage their natural spatial reasoning and memory~\cite{lisle_sensemaking_2021, skarbez_immersive_2019, yousif_using_2021}. 
Users can physically navigate around and within the graph, examining connections from multiple perspectives. This embodied navigation transforms the virtual space into an external cognitive aid~\cite{andrews2011information, hutchins1995cognition, kirsh1995intelligent,yang2020embodied,reiske2023multi}, potentially making complex relationships easier to comprehend compared to flat representations. 
This spatial layout forms the foundation for all subsequent foraging and sensemaking activities within \tool{}.


\subsection{Multimodal Interaction Model}


To ensure users can control the system easily and support transitioning between finding and analyzing literature, \tool{} uses a combination of interaction methods: direct hand gestures, voice commands, and a WIMP-style hand-menu interface~(\ref{goal:interaction}). Offering multiple input modes is crucial in complex VR environments because it provides flexibility; users can choose the most efficient or comfortable method for a given task, reducing interaction friction and mental load~\cite{kim_datahand_2021, multimodal_microsoft, saktheeswaran_touch_2020}.



\smallskip
\noindent
\textbf{Voice Commands.} For actions like requesting summaries (``\textit{Summarize paper}'') or triggering recommendations (``\textit{Recommend papers by thematic similarities}''), voice input offers a fast, hands-free alternative. This modality is particularly beneficial when users' hands are occupied with manipulation tasks or when navigating complex menu structures would be cumbersome~\cite{cohen1995role}. Speech allows for efficient execution of commands and queries using natural language, reducing the need for users to divert visual attention or exert physical effort, thus supporting fluid interaction~(\ref{goal:interaction}).


\smallskip
\noindent
\textbf{Embodied Gesture Interaction.}
Users interact directly with paper nodes using hand tracking. A simple pinch gesture selects a paper, which can then be moved freely in 3D space (6DoF) and pinned. This leverages the principles of direct manipulation~\cite{shneiderman1983direct}, making spatial organization tasks feel intuitive and mapping directly to users' actions. 
Such embodied interactions are particularly powerful in VR, allowing users to leverage their innate spatial reasoning and proprioception, effectively turning the virtual space into an extension of their cognition~\cite{dourish2001action,laviola20173d,kirsh1995intelligent}.
Gestures like bringing two papers together to link them (bimanual interaction)~(\autoref{fig:gesture}\figpart{A}) or the steepling gesture for clustering~(\autoref{fig:gesture}\figpart{B})
further enhance this embodied control, aligning with findings that physical actions can offload cognitive burden in immersive analytics~\cite{cordeil_imaxes_2017,in_this_2024,yang_tilt_2021}.
    

\setlength{\columnsep}{5pt}%
\setlength{\intextsep}{0pt}%
\begin{wrapfigure}{R}{0.27\textwidth}
   \vspace{0pt}
   \centering
   \includegraphics[width=0.27\textwidth]{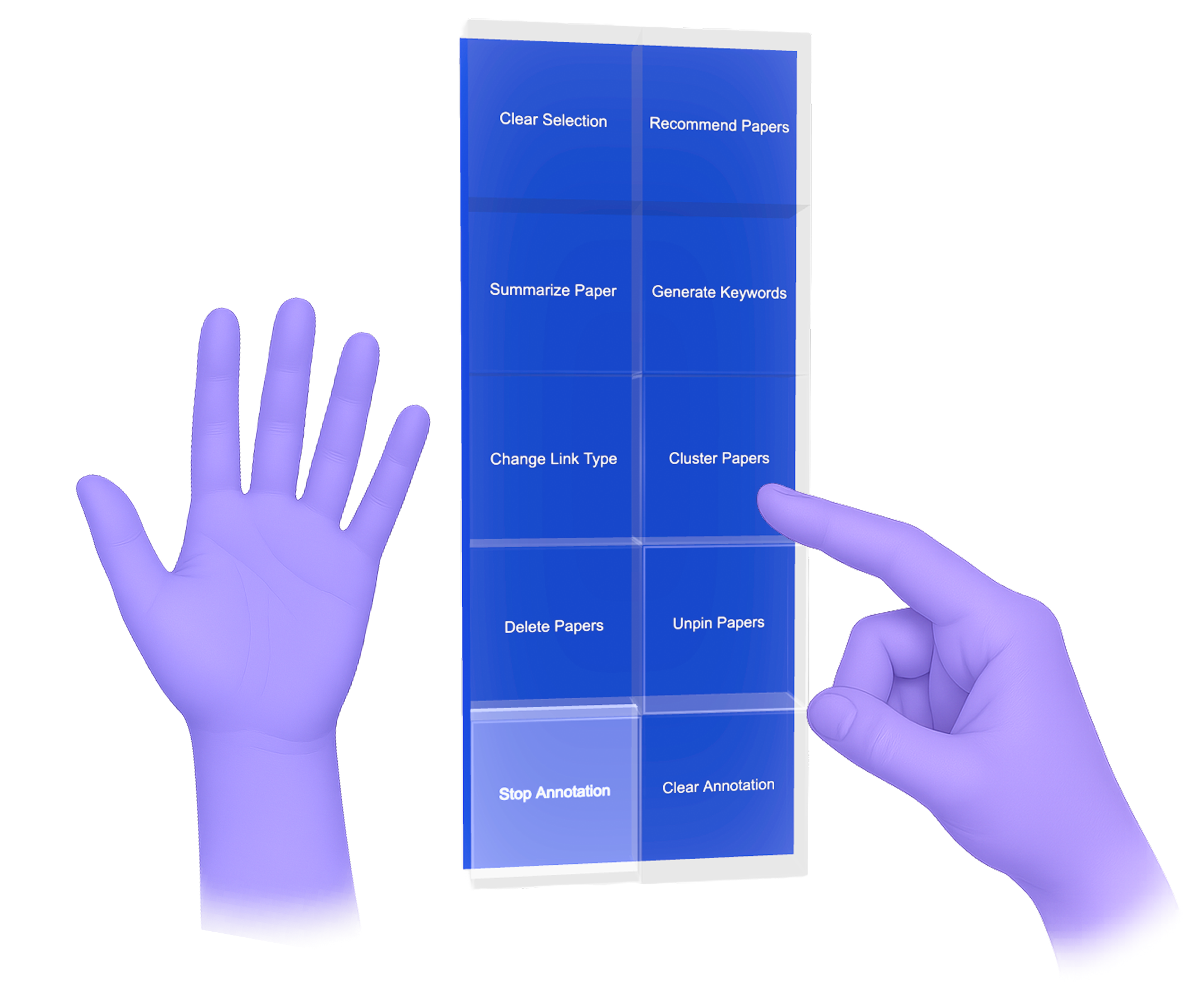}
   \label{fig:hand-menu}
   \vspace{-10pt}
\end{wrapfigure}

\smallskip
\noindent
\textbf{Hand Menu.} A standard menu attached to the user’s non-dominant hand provides reliable, visible access to all features using familiar point-and-click interaction with holographic buttons. This WIMP-style interface acts as a dependable fallback, enhances the discoverability of system functions~\cite{norman2013design}, and ensures usability for users who may prefer explicit control, are new to voice/gesture commands, or encounter limitations with other modalities. Such menus remain a cornerstone of VR interaction design for providing structured access to functionality~\cite{laviola20173d}.





\subsection{Intelligent Foraging}
\label{4.3}


\tool{} includes several features designed specifically to help with finding, understanding, and organizing literature.

\subsubsection{Intelligent Network Expansion via Recommendations}
\tool{} offers multiple ways to expand the literature network, helping users discover relevant papers as they forage~(\ref{goal:guidance}). 
This is primarily achieved through an intelligent paper recommendation system. 
Users can request recommendations via the hand-menu or corresponding voice commands (e.g., ``\textit{Recommend papers by citations}''). 
The system offers three key recommendation modes---\textit{thematic similarities}, \textit{citations \& references}, and \textit{author-centric}, reflecting common literature exploration strategies and corresponding to the literature network's visualization types~(\autoref{4.1}).
These modes respectively capture content-based, structural, historical, and community-driven discovery, providing a comprehensive yet manageable toolkit for navigating the literature landscape.

\smallskip
\noindent\textbf{Thematic Similarities.}
Users can request recommendations based on thematic similarities to one or more selected papers, enabling the addition of topically relevant yet potentially novel works to the current network view. 
A powerful application of this is creating thematic ``bridges'': by selecting papers from seemingly unrelated clusters and requesting recommendations, users can uncover non-obvious connections between research areas, surface novel trends, and receive guided suggestions for exploration.

\smallskip
\noindent\textbf{Citations \& References.}
Users can expand the network by exploring citation links. When a single paper is selected, it can retrieve its ``forward citations'' (papers citing the selected work) or ``backward references'' (papers cited by the selected work). 
This common strategy allows researchers to trace the lineage and impact of ideas over time, providing insights into a field's development and trajectory.

\smallskip
\noindent\textbf{Author-Centric.}
Users can explore an author's papers by selecting their name on a paper node to request other publications by that author. This helps researchers identify key experts within specific areas and discover relevant work through established authors.

\begin{figure}
    \centering
    \includegraphics[width=0.85\linewidth]{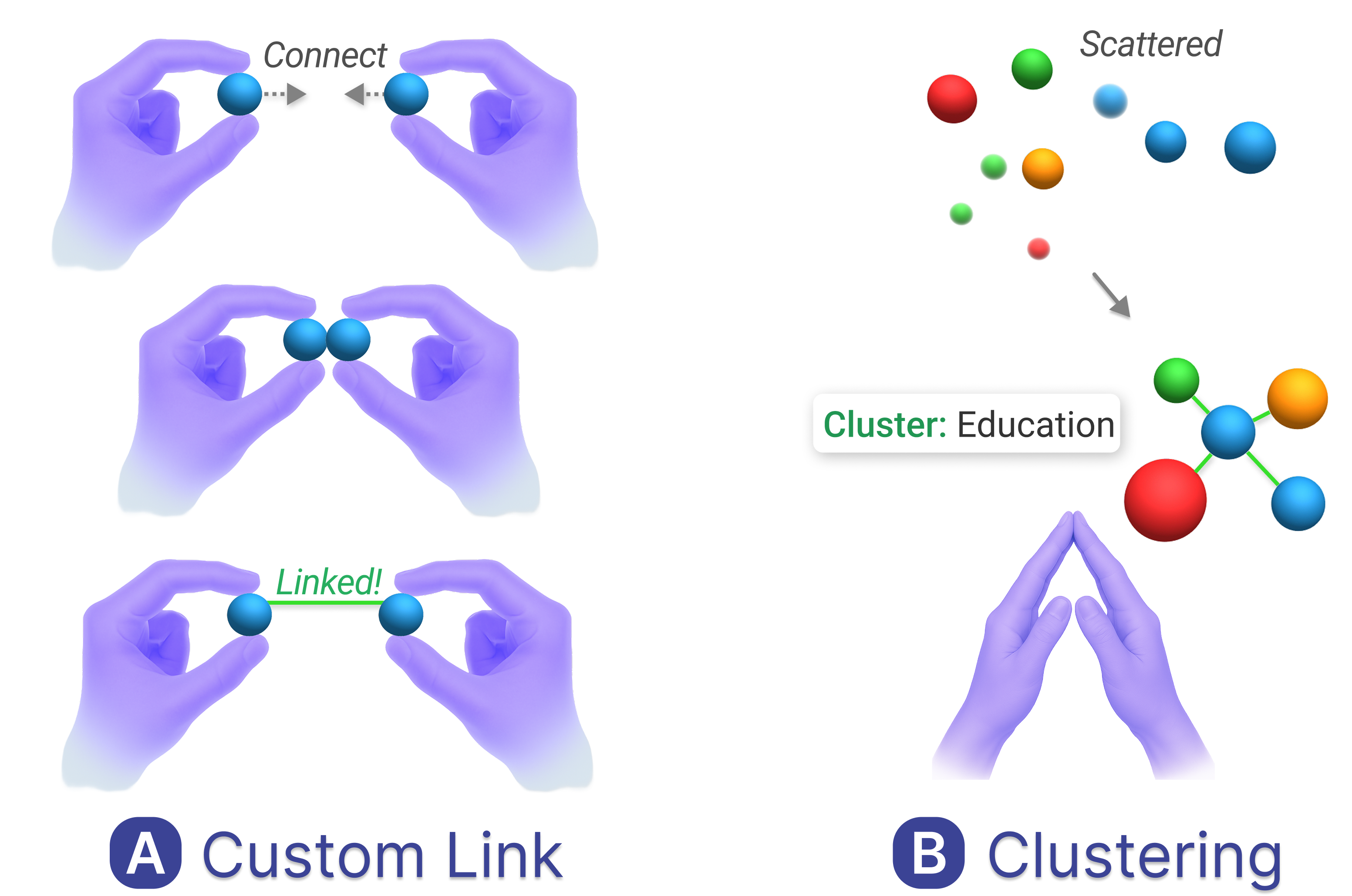}
    \caption{Two bimanual gestures for organizing papers in \tool{}. \textbf{(A)} Bringing two papers together with a bimanual pinch gesture creates a \textbf{Custom Link}. \textbf{(B)} A bimanual steepling gesture triggers automatic \textbf{Clustering} of nearby papers.}
    \label{fig:gesture}
\end{figure}

\subsubsection{AI-Enabled Thematic Clustering and Custom Links}
\label{4.3.2}

To support flexible organization within literature networks, \tool{} provides both automated and manual structuring tools for users to personalize their literature network~(\ref{goal:organization}). These features help reduce cognitive load while enabling researchers to map connections, identify patterns, and externalize their evolving understanding of the research landscape.

\smallskip
\noindent\textbf{LLM-enabled Topic Clustering.}
\tool{} uses Google’s Gemini 2.5 Flash~\cite{Gemini2.5Flash} to automatically group papers into thematic clusters based on title and abstract analysis.
These clusters are spatially arranged and labeled with their core themes, effectively categorizing the network into key domains. 
This visualization acts as a powerful guide for exploration and enables efficient spatial organization and thematic externalization~(\ref{goal:guidance}, \ref{goal:organization}). 
The clustering feature is readily accessible via a hand-menu button, the voice command (\textit{``Cluster papers''}), or the bimanual steepling gesture~(\autoref{fig:gesture}\figpart{B}).

\smallskip
\noindent\textbf{Customized Node Links.}
Beyond automatically generated connections, users can manually create custom links between nodes. By physically bringing two nodes close together (one held in each hand), users establish a direct link. This allows researchers to define unique relationships that may not be captured by citation data or recommendation algorithms, enabling them to flexibly structure the network according to their personal conceptual models and spatially externalize their understanding (\ref{goal:organization}).

\subsubsection{Intelligent Paper Insights and Annotation}

To help users navigate information overload~\cite{landhuis_scientific_2016} and quickly assess relevance, \tool{} integrates AI-powered analysis based on each paper’s metadata via Google’s Gemini 2.5 Flash API~\cite{Gemini2.5Flash}. To support synthesis, it also offers lightweight annotation tools.

\begin{figure}
    \centering
    \includegraphics[width=\linewidth]{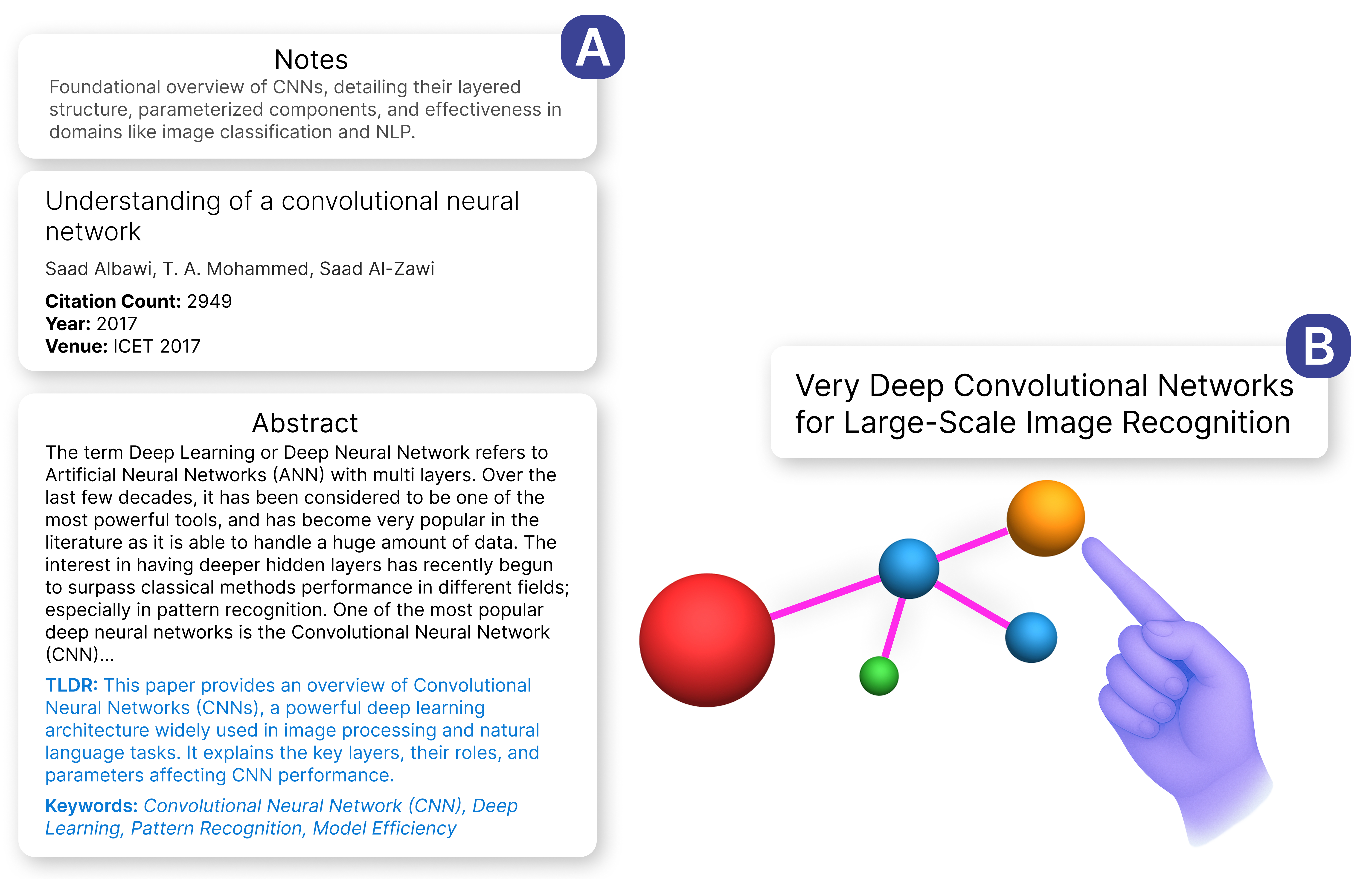}
    \caption{User’s point of view of the paper assessment interface. \textbf{(A)} \textbf{Paper Insights Panel} with detailed metadata (e.g., authors, venue), AI insights including TLDR and keywords, and user-authored annotations. \textbf{(B)} Hovering over a paper node reveals its title above the node for quick identification without opening the full panel.}
    \label{fig:paper-panel}
\end{figure}

\smallskip
\noindent
\textbf{Paper Insights Panel.}
Hovering over a node reveals its title for quick scanning~(\autoref{fig:paper-panel}\figpart{B}), while long pressing opens a detailed panel with the title, authors, and abstract of the paper~(\autoref{fig:paper-panel}\figpart{A}). The panel remains pinned near the node and oriented toward the user to preserve readability and spatial context. These interactions provide quick access to essential information, reducing cognitive load and supporting fluid navigation~(\ref{goal:guidance}, \ref{goal:interaction}).


\smallskip
\noindent
\textbf{AI-Driven Content Analysis.}
From the panel, users can request a quick summary (“TLDR”) or keywords via menu or voice commands~(\autoref{fig:paper-panel}\figpart{A}). Powered by LLM, these features help distill key information, reducing the effort and time needed to assess a paper’s relevance~(\ref{goal:interaction}). 

\smallskip
\noindent
\textbf{Speech-Based Annotations.}
Users can dictate notes onto a paper via a hand-menu button or the voice command (``\textit{Start/stop annotating}''). The system transcribes the speech and attaches the notes to the panel~(\autoref{fig:paper-panel}\figpart{A}). This allows researchers to capture immediate thoughts without breaking their exploration flow, smoothly bridging foraging and synthesis~(\ref{goal:interaction}), which aligns with sensemaking models emphasizing iterative reflection.

\section{User Study}
\label{5}
%

\begin{figure*}
    \centering
    \includegraphics[width=\linewidth]{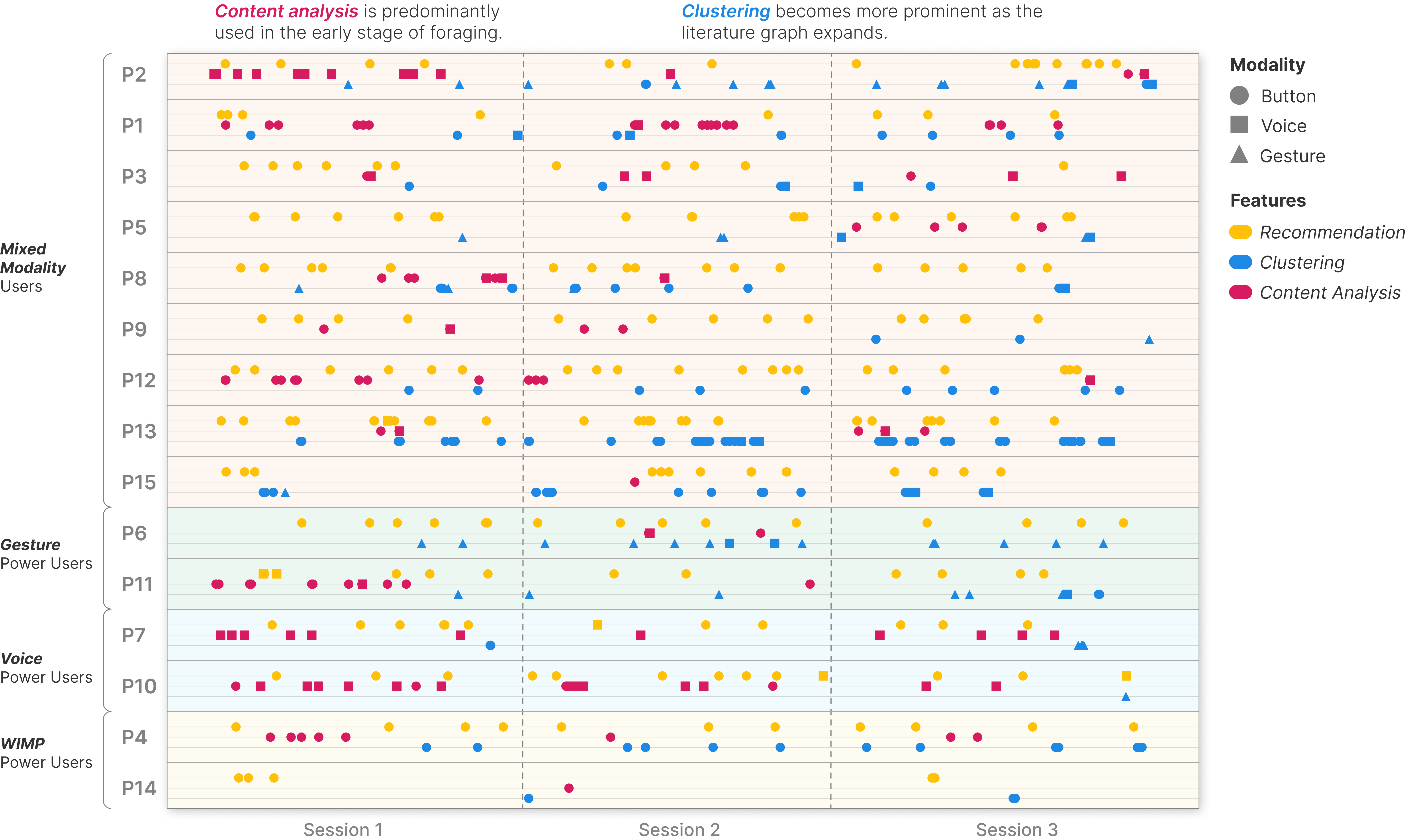}
    \caption{Interaction logs of all participants across three sessions, showing input modalities (button, voice, gesture) and feature usage (recommendation, clustering, content analysis). Users exhibited diverse modality preferences, including mixed-modality users and modality-specific  (gesture, voice, WIMP) power users.
}
    \label{fig:interaction-logs}
    \vspace{-3pt}
\end{figure*}

\added{We conducted an observational study to understand how researchers utilize immersive tools for literature exploration and synthesis, focusing on their strategies, behaviors, and preferences. Rather than measuring task performance, our qualitative emphasis reflects \tool{}’s aim to support open-ended sensemaking and research ideation. The study scenario was informed by insights from our formative expert interviews and grounded in common document foraging tasks described in literature review best practices. This alignment ensured that the scenario realistically exercised the system’s core features and provided comprehensive coverage of relevant user workflows.}
Accordingly, we formulated the following research questions:

\begin{enumerate}[topsep=2pt, itemsep=0mm, parsep=3pt, leftmargin=26pt, wide, labelwidth=!, labelindent=0pt, label=\textbf{RQ\arabic*:}, ref=G\arabic*]

    \item \textit{How do users choose between different input modalities (gesture, speech, or hand-based menu) when managing the exploration and organization process?}
    
   \noindent We explore whether modality preferences are task-specific, user-dependent, or influenced by system affordances, and how input choice affects user flow, agency, and control in immersive interaction.

    \item \textit{How do users externalize their understanding by organizing, annotating, or clustering papers within the 3D immersive space?}
    
    \noindent We seek to understand the diversity of spatial structuring and annotation strategies users employ, and how these externalization practices contribute to managing complexity and potentially offloading cognitive load during immersive literature foraging.

    \item \textit{How do users develop and refine their sense of relevance during immersive literature exploration to guide subsequent foraging actions?}
    
    \noindent We aim to understand the process by which users determine paper relevance, identify the factors influencing these judgments (e.g., metadata, network cues, task goals), and analyze how their relevance criteria evolve throughout the exploration session.

\end{enumerate}

\smallskip
\noindent\textbf{Participants and System Setup.}
We recruited 15 participants \added{(P1–P15; 8 female, 7 male; ages 20–27)}, including 9 PhD students, 5 master’s students, and 1 undergraduate student, with backgrounds in information visualization, HCI, XR, machine learning, and education. Most had 1–4 years of research experience and regularly engaged in literature reviews, 
with 10 identifying as experts, and five as occasional. 
All but one had used reference managers, and several had tried academic visualization or AI-based tools. Most had little or no prior experience with immersive technologies. 

The experiment was conducted within a 10$\times$10 ft lab area, ensuring sufficient space for participants to move freely. 
A Meta Quest 3 headset was used. Participants interacted with the \tool{} using the headset's built-in hand tracking and voice commands facilitated by an external microphone. 
The experimenters observed the session by monitoring the participant's VR view cast to a laptop and their physical movements.

\smallskip
\noindent\textbf{Task Scenario and Goals.}
The study used a task scenario simulating the initial phase of a literature review. Participants were asked to construct a \textit{Related Work} graph for a research topic relevant to their background. 
This task was designed to mirror the early stages of scholarly inquiry, where researchers identify, evaluate, and organize literature to understand a research landscape. Each participant received a topic aligned with their academic background. 
To initiate exploration, three seed papers representing distinct sub-topics were preloaded into \tool{}. 
These papers served as the starting point. Participants had full autonomy in deciding which papers to investigate further or expand upon. Participants were instructed to explore the literature using the system's recommendation features, assess the relevance of discovered papers, and arrange them spatially. 
We intentionally avoided prescribing a specific structure or output, encouraging participants to arrange papers in a personally meaningful way that externalized their reasoning and sensemaking approach.

\smallskip
\noindent\textbf{Experimental Procedure.}
Each study session lasted approximately 90 minutes and consisted of four parts: pre-study briefing, system tutorial, main exploration task, and post-task debriefing. The session began with a 5-minute pre-study phase for consent, a brief welcome, and a short demographic questionnaire.

Next, participants proceeded to a 30-minute interactive tutorial. This extended duration was deemed necessary to ensure that participants achieved proficiency with the system's multiple interaction modalities (e.g., hand tracking, voice commands) and core literature exploration features. Wearing the VR headset, they familiarized themselves with these functions under the real-time guidance of the researcher observing via the cast view. They were encouraged to ask questions throughout the walkthrough and were given time to practice interactions at their own pace afterwards.
The main exploration task lasted 40 minutes, divided into three 10-minute sessions with short breaks in between. This structure helped mitigate potential VR fatigue while enabling us to observe how participants’ strategies evolved over time. At the end of each session, we conducted a brief 1-minute interview in which participants summarized their goals, progress, and plans for the next phase.\added{ These periodic reflections served as a lightweight alternative to traditional think-aloud protocols, capturing participants’ reasoning without disrupting task focus or interfering with voice-based interactions.} Each summary was followed by a 4-minute rest break, during which participants could relax or ask questions.

Following the main task, a 15-minute semi-structured interview session was conducted to gather rich qualitative feedback on their experience and probe specific observed behaviors.

\smallskip
\noindent\textbf{Data Collection and Analysis.}
\added{We collected behavioral data through interaction logs, brief in-task reflections, and a post-task semi-structured interview.}

\noindent 
\textit{Interaction Logs.}  
The system automatically recorded timestamped user interactions, capturing exploration sequences (e.g., paper selections, recommendation requests), spatial manipulations (e.g., moving, clustering papers), and input modality usage (gesture, speech, WIMP menu). These detailed behavioral traces enable the reconstruction and analysis of users' navigation paths and interaction patterns throughout the session.

\noindent 
\textit{Self-Reflections and Interviews.}  
During short breaks between sessions, participants provided brief verbal reflections, articulating their goals, progress, and next steps. These reflections offered insight into participants’ evolving strategies in real time. After the main task, we conducted a semi-structured interview to further explore their experiences, including their relevance judgment criteria, spatial structuring rationale, and input modality preferences.



\section{Results \& Discussion}
\label{6}
\added{Our analysis aims to identify distinct information foraging patterns in the immersive environment. We incorporated both quantitative and qualitative data in our analysis: interaction logs and Likert-scale responses were examined to identify behavioral trends, while interviews and observational notes were analyzed using thematic analysis.}
In this section, we discuss how immersive affordances affect iterative foraging and synthesis behaviors.

\subsection{Input Modality Preferences and Tradeoffs (RQ1)}
To explore RQ1, we analyzed all 15 participants’ interaction logs within our immersive literature exploration system during the user study. This analysis aims not only to identify modality usage patterns, but also to understand the underlying strategies, tradeoffs, and implications for usability, expressiveness, and cognitive effort in immersive environments. We structured our analysis along two key dimensions: (1) Input Modality---\textbf{gesture}, \textbf{speech}, or \textbf{hand menu} (WIMP-style input); and (2) System Features---\textit{Recommendation}, \textit{Clustering} (system-generated topic clusters and customized node links), and \textit{Content Analysis} (AI-based summarization, keyword extraction, and user-created annotations). Visualizing modality and feature usage across sessions revealed several key patterns~(\autoref{fig:interaction-logs}).


\smallskip
\noindent\textbf{WIMP Interface as the Ubiquitous Default.}
Across all foraging categories, \textbf{hand menu} emerged as the most frequently used modality. This was especially pronounced for \textit{Recommendation}, where every participant relied almost exclusively on button input over speech. This strong reliance on WIMP-style menus reflects users' comfort with explicit, visible UI controls---particularly for discrete, function-oriented tasks like querying for related papers. 

\smallskip
\noindent\textbf{Mixed Modality Use as a Marker of Adaptivity and Expertise.}
Most participants (9/15) exhibited active mixed modality usage---combining at least two input modalities over the course of the sessions. Typically, users began with \textbf{hand menu} interactions, gradually incorporating gestures or voice as they gained comfort and familiarity. Among these mixed modality users, P2 stood out as particularly adaptive and fluent. P2 not only explored all three modalities but also demonstrated a high degree of contextual sensitivity in modality choice---using \textbf{gesture} for \textit{Clustering}, \textbf{voice} for \textit{Content Analysis}, and \textbf{hand menu} for other precision or unfamiliar actions. We also observed several \textit{power users} who demonstrated proficiency in specific modality-feature pairings. For clustering tasks, P6 and P11 consistently leveraged the bimanual steepling gesture to invoke clustering. For content analysis, P7 and P10 became strong voice users, frequently invoking content analysis through speech. 
This flexible, situational use of input demonstrates a desirable interaction pattern for immersive systems, where users fluidly combine multiple modalities to best suit their preferences, context, and task.

\smallskip
\noindent\textbf{Foraging Strategies Blended Recommendation-Driven and Spatial Exploration.}
Our analysis revealed an important evolution in participants' foraging strategies. 
Initially, users relied predominantly on system guidance via \textit{Recommendation} and \textit{Content Analysis}. As their spatial understanding of the literature collection grew, their activity shifted towards self-directed exploration using \textit{Clustering} and manipulating the layout. 
This demonstrates a natural blending of strategies, leveraging both system guidance and the emerging spatial context. 
This underscores the need for immersive systems to seamlessly integrate recommendation features within the spatial exploration environment.

\smallskip
\noindent\textbf{Summary.} 
Our findings show participants initially relied on hand menus for precise tasks like selecting recommendations, reflecting a preference for familiar, structured interactions that reduce errors~\cite{norman2013design}.
As users became more comfortable, many started mixing input methods, using gestures and voice for actions where those modalities felt more natural or expressive.
This reflects how users often combine different inputs to leverage their unique strengths~\cite{bolt1980put,oviatt1999ten}.
Importantly, even with novel VR inputs available, visible menus remained essential support, especially for complex tasks or new users, highlighting the need for clear guidance in 3D environments~\cite{laviola20173d}.
Therefore, future immersive systems should offer a balanced mix of input options, tailored to the specific task.

\begin{figure}
    \centering
    \includegraphics[width=\linewidth]{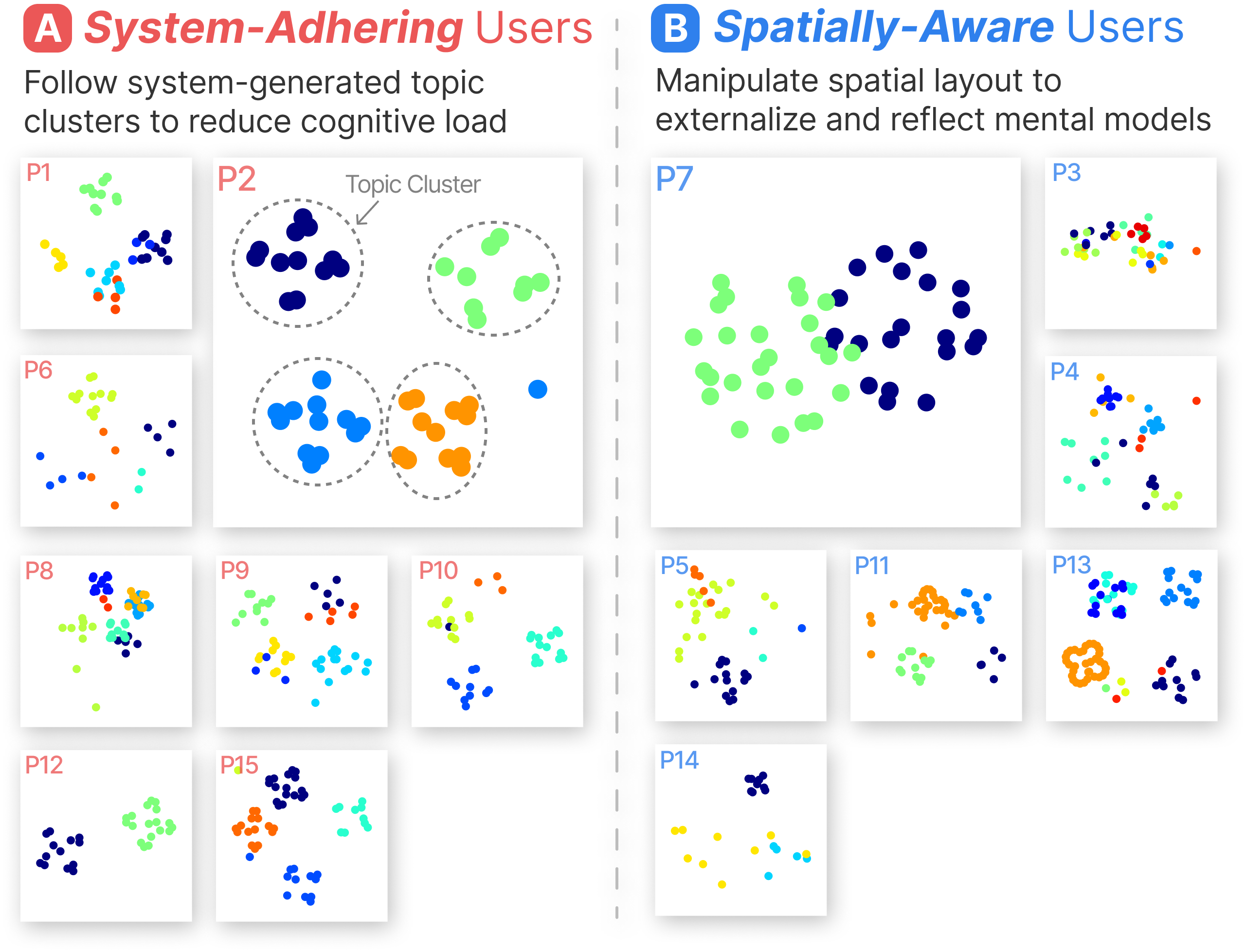}
    \caption{Final spatial layouts of participants’ literature networks (bird’s-eye view), illustrating two distinct spatial organization behaviors. Colors indicate system-generated clusters.
    \textbf{(A) System-Adhering Users} (left, e.g., P2) largely preserved system-defined clusters, while 
    \textbf{(B) Spatially-Aware Users} (right, e.g., P7) actively reorganized papers to reflect their own mental models. }
    \label{fig:final-2d-layout}
    \vspace{-3pt}
\end{figure}

\subsection{Spatial Externalization Strategies (RQ2)}
To address RQ2, we analyzed participants' spatial organization behaviors throughout their literature exploration sessions. \added{All participants demonstrated meaningful and sometimes complex use of spatial manipulation to externalize their evolving understanding of the literature network in immersive space.} By examining both the participants’ final configurations and temporal progression of layouts, we identified key strategies used to support sensemaking.

\smallskip
\noindent\textbf{Clustering and Spatial Grouping Behavior.}
A consistent pattern observed across almost all participants was the use of clustering to organize papers spatially. 
\added{Most participants’ final layouts exhibited at least one or more clearly defined, visually distinct clusters~(\autoref{fig:final-2d-layout}), often aligned with the system-generated thematic clusters (color-coded).
This suggests that automated clustering served as a helpful starting point for organization and, in many cases, as a foundation for participants’ externalization strategies.}

\added{We observed a spectrum of spatial behaviors that reflected how closely participants' mental models aligned with the system-generated structure.
We distinguish between \textbf{System-Adhering} and \textbf{Spatially-Aware} users. \textbf{System-Adhering} users (e.g., P2) relied heavily on the automatic clustering with minimal manual reorganization~(\autoref{fig:final-2d-layout}\figpart{A}); 8 out of 15 participants followed this approach. This suggests that the LLM-based clustering often captured meaningful thematic boundaries sufficient for users’ foraging needs.}

\added{In contrast, \textbf{Spatially-Aware} users (e.g., P5, P7, P11, P13) employed more nuanced and active spatial strategies~(\autoref{fig:final-2d-layout}\figpart{B}). These participants merged, split, or repositioned clusters based on personal interpretations of topic relevance and research goals. For instance, P7 described their process as “creating a spider web [of papers],” deliberately emphasizing central topics and constructing “meta-clusters” to indicate relationships between system-defined themes. Some also used spatial properties to encode importance or intent (e.g., placing key papers in direct view and moving less relevant ones behind them). These behaviors underscore how embodied spatial interaction in VR enabled personalized and expressive externalization beyond static system outputs.}

\begin{table*}
\small
\centering
\caption{Summary of relevance-judgment themes for RQ3 (N=15 participants).}
\label{tab:rq3_summary}
\begin{tabular}{p{5cm} p{1.2cm} p{10cm}}
\toprule
\textbf{Theme} & \textbf{n/15} & \textbf{Representative Quote} \\
\midrule
Thematic similarities as primary channel & 15 & “I relied mostly on thematic similarities because I wanted papers with similar context.” (P1) \\
Citation / reference trails for bridging & 10 & “I relied heavily on citation and reference recommendations, then used thematic to broaden.” (P10) \\
Pivoting among recommendation modes & 9 & “At first I relied on references, then switched to thematic when I needed new ideas.” (P2) \\
Metadata scanning (title, keywords, summary) & 7 & “Due to time limits, I mostly relied on titles; if one seemed off-topic, I excluded it.” (P10) \\
Author-based recs seldom used & 5 & “I didn’t use author recs at all — I’m not familiar with authors in this field.” (P5) \\
Ad-hoc → patterned workflows & 5 & “I started kind of sporadically, but then developed a pattern as I went along.” (P1) \\
Graph-based heuristics (centrality / clusters) & 4 & “I selected papers based on how closely they were connected to others in the network.” (P3) \\
Quantity vs. precision shifts & 4 & “Initially I was picky, but later I realized I’d just add anything that might be relevant.” (P2) \\
Citation maturity checks & 3 & “A lot of recent papers with zero citations felt odd; I preferred older, foundational work.” (P8) \\
Personal interest override & 3 & “I decided relevance mostly by personal interest, especially around XR in education.” (P7) \\
Stable approach (little or no change) & 1 & “My core method didn’t really change through the sessions.” (P7) \\

\bottomrule
\end{tabular}
\vspace{-1.5em}
\end{table*}

\begin{figure}
    \centering
    \includegraphics[width=\linewidth]{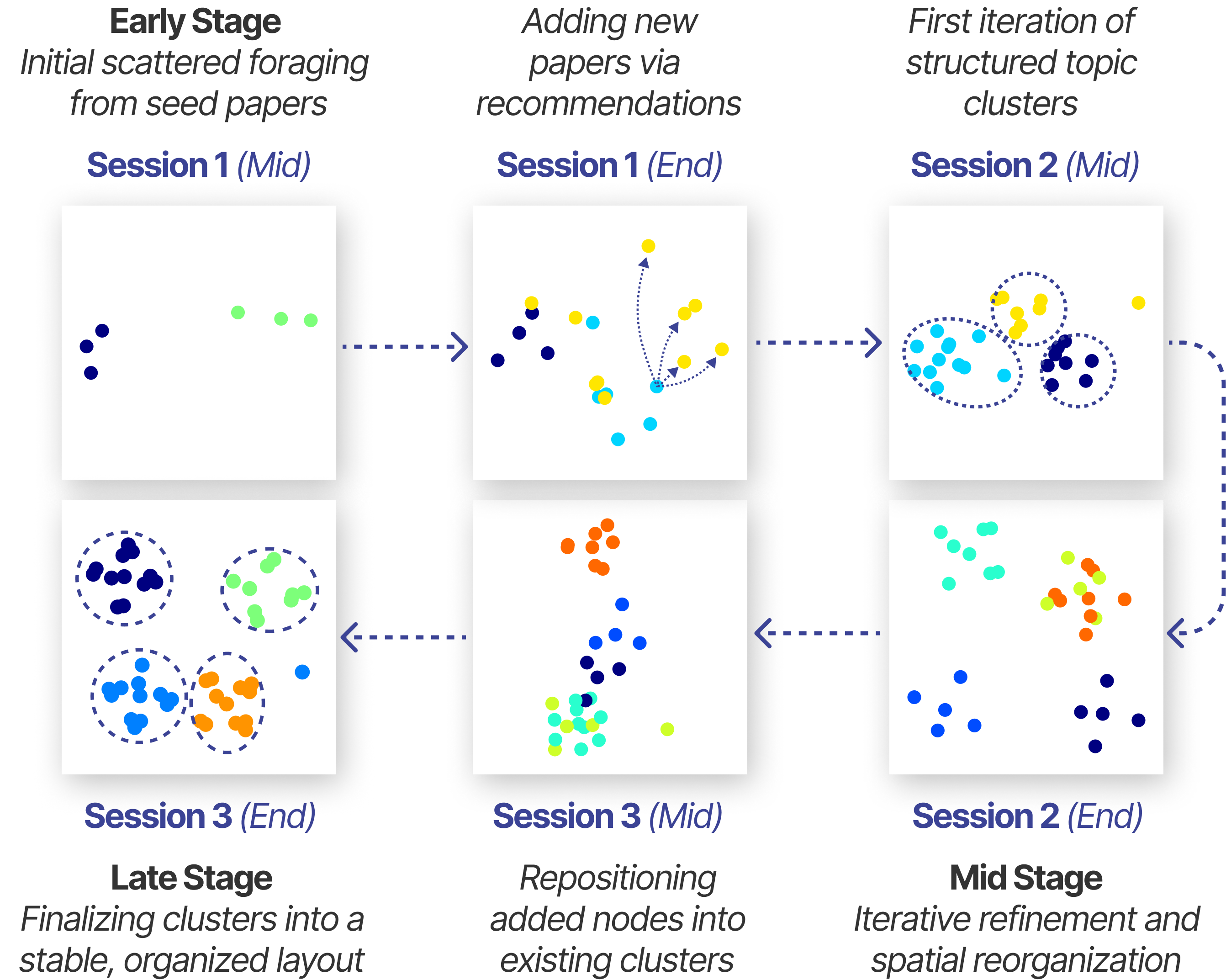}
    \caption{Temporal evolution of Participant 2’s spatial layout, showing six snapshots of the literature network from a bird’s-eye view. \textit{Mid} and \textit{End} indicate the middle and end of each session, respectively.
    }
    \label{fig:layout-over-time}
     \vspace{-3pt}
\end{figure}

\smallskip
\noindent\textbf{Evolution of Spatial Organization Over Time.}
Beyond analyzing final layouts, we also analyzed the temporal evolution of participants' spatial layouts to understand their sensemaking process (\autoref{fig:layout-over-time} shows P2's progression). Initially, layouts tended to be sparse, reflecting early exploration based on seed papers and system recommendations. However, as participants gathered more literature, they began actively structuring their space—repositioning items, creating thematic clusters, and refining the overall organization. Final layouts were typically more structured and clustered, visually representing the user's developed understanding. This progression demonstrates that spatial externalization was not fixed but emerged dynamically through the user's iterative cycle of foraging, organizing, and understanding. Clustering stood out as an intuitive method for managing the growing complexity, helping users reduce clutter while maintaining important thematic connections.

\smallskip
\noindent\textbf{Summary.}
Overall, these findings demonstrate that spatial externalization was a core strategy in participants’ immersive literature foraging. 
Participants actively engaged in clustering, iterative reorganization, and personalized layout construction to manage complexity and externalize their understanding within the 3D environment.
This use of space to aid thinking is a known cognitive strategy~\cite{hutchins1995cognition,kirsh1995intelligent} and matches behaviors seen in similar VR~\cite{lisle_sensemaking_2021} and large-screen~\cite{andrews2011information} sensemaking tools.
The combination of automated clustering and manual spatial manipulation enabled users to create meaningful, interpretable representations of the literature network.

\subsection{Forming and Adjusting Relevance (RQ3)}
This section examines how participants judged paper relevance, what shaped those judgments, and how their strategies evolved over time. \added{Our thematic analysis~\cite{braun_reflecting_2019, byrne_worked_2022} was collaboratively conducted by two authors to identify recurring patterns.} As summarized in~\autoref{tab:rq3_summary}, users engaged in a layered sensemaking process that combined familiar metadata heuristics with network structure and system recommendations.

\smallskip \noindent\textbf{How Relevance Was Judged.}
Most participants began with fast filtering based on visible metadata—especially paper titles. Seven out of 15 participants made keep-or-discard decisions primarily from title scans, often skipping abstracts. A few also referenced prompt-related keywords or short summaries. A subset (3/15) checked citation counts to assess paper maturity, preferring well-cited works. Three participants admitted that personal interest influenced their choices, sometimes retaining papers only loosely related to their search goals. Together, these strategies reflect a common filtering process: participants typically began with a quick assessment based on easily visible cues (like titles or keywords), then applied more careful filtering using abstracts or citation counts when needed, and occasionally explored papers out of personal curiosity. Across participants, these heuristics prioritized efficiency and ease of use, while still allowing flexibility to follow individual interests and goals during immersive literature foraging.

\smallskip \noindent\textbf{Factors Influencing Relevance Judgments.}
Recommendation tools strongly shaped exploration patterns. All participants used recommendations by thematic similarities for topical consistency and interdisciplinary insights.  Two-thirds of participants (10/15) also used citation or reference views to discover foundational papers or bridge clusters. Author-based recommendations were rarely used (5/15), mostly due to unfamiliarity with domain experts. Spatial layout also influenced judgments. Four out of 15 participants referenced cluster centrality or connectivity, using network position as a proxy for importance when metadata was insufficient.

\smallskip \noindent\textbf{How Strategies Evolved Over Time.}
Most participants developed dynamic adaptive strategies throughout the study. Five participants began with ad-hoc browsing before settling into structured workflows. The majority (9/15) reported switching recommendation modes---for example, moving from citation-based to thematic recommendations---when their progress slowed or when trying to seek different types of connections. Some (4/15) adjusted their level of selectivity over time, shifting between quantity-first (gather broadly) and precision-first (filter narrowly) approaches. Two participants even developed new goals mid-task, such as connecting previously separate clusters of papers. Notably, only one participant maintained a consistent strategy throughout the session. This adaptability underscores the inherently serendipitous nature of literature exploration---not merely a process of retrieving known items, but an evolving search for connections, gaps, and unexpected insights within a complex and expanding information space.

\smallskip \noindent\textbf{Summary.}
Across sessions, participants combined filtering heuristics with recommendation tools and network cues to refine their relevance judgments. Participants also exhibited distinct foraging styles---such as \textit{metadata scanners} (filtering by titles or keywords), \textit{network navigators} (navigating the graph structure), and \textit{adaptive pivots} (frequently switching strategies).
These styles suggest opportunities for future personalization in immersive systems. More broadly, these patterns highlight how immersive spatial tools actively shape how users assess and construct relevance during open-ended, exploratory tasks like literature review.

\section{Conclusion \& Future Work}
We present \tool{}, an immersive analytics system designed to facilitate the integrated workflow of literature exploration. Through a formative study with academic researchers, we identified essential user needs and established key design goals: exploration guidance, spatial organization, and seamless transitions between foraging and synthesis. \tool{} achieves these goals by combining interactive 3D literature network visualization, flexible spatial organization, and multimodal interactions. Our user study with 15 researchers revealed distinct patterns in modality preference, spatial organization, and relevance formation. This work highlights the potential of immersive technologies to enhance literature foraging and contributes to immersive analytics by addressing previously underexplored aspects of the sensemaking workflow.

Future work will explore enhancing \tool{} by integrating personalized recommendation algorithms and expanding multimodal interaction capabilities. Additionally, longitudinal studies will investigate how sustained use impacts researchers' productivity and cognitive workflows in real-world scenarios.


\acknowledgments{
This work was supported in part by NSF grant IIS-2441310.
}

\bibliographystyle{abbrv-doi-hyperref}

\bibliography{references}

\end{document}